\documentclass[11pt]{article}
\usepackage[utf8]{inputenc}
\usepackage[T1]{fontenc}
\usepackage{amsmath,amssymb}
\usepackage{graphicx}
\usepackage[margin=1in]{geometry}
\usepackage{hyperref}

\title{Reveal normal form structure for nonlinear map in accelerator beam physics with symplectic neural network}

\author{Leo He$^{1}\thanks{
Email: lhe@stu.3villagecsd.org or leohe1310@gmail.com.
This project was supported in part by the Brookhaven Science Associates(BSA) under the BNL High School Research Program(HSRP).
}$ \and  Yue Hao$^{2,3}$ }

\date{}

\usepackage{tikz}
\usetikzlibrary{positioning,arrows.meta}
\begin{document}

\maketitle

\begin{center}
\footnotesize
$^{1}$Ward Melville High School, NY 11733\\
$^{2}$Department of Physics and Astronomy, Michigan State University, MI 48824\\
$^{3}$Collider Accelerator Department, Brookhaven National Lab, NY 11973
\end{center}

\vspace{1em}

\begin{abstract}

Long-term multi-turn tracking is required to study nonlinear phenomena
in circular accelerators, such as resonances, chaotic motion, and the
dynamic aperture. General-purpose neural networks do not preserve the
symplectic structure of Hamiltonian dynamics, so their errors can grow
over many turns. Symplectic neural networks (SympNets) guarantee
symplecticity through their architecture rather than through a loss
penalty. In this work, we use a SympNet to learn the normal-form
transformation of a nonlinear map from tracking data. In the learned
coordinates, the dynamics become a rotation of each mode, whose
amplitude-dependent phase advance is given by a second network. We demonstrate the method on a four-dimensional
McMillan-type map. The model reproduces the one-turn dynamics, the learned
phase advances stay constant along each trajectory to within
$\sim 10^{-5}$~rad, and orbits become close to circles in the learned
coordinates at small and moderate amplitude. At large amplitude the
learned orbits spread noticeably, and the cause of this degradation is
not yet established. The approach is a step toward fast,
structure-preserving surrogate models for lattice analysis and online
beam-dynamics applications in storage rings.

\end{abstract}

\section{Introduction}

Nonlinear fields from sextupole and octupole magnets and from lattice
imperfections limit the dynamic aperture of modern circular
accelerators. They make the betatron tune depend on amplitude
(amplitude-dependent tune shift, ADTS), distort the invariant curves
in phase space, and can drive particles onto resonances, with
consequences for beam lifetime and injection efficiency
\cite{chao2013handbook}. Accurate models of these amplitude-dependent
nonlinearities are therefore essential for accelerator simulation and
stability analysis. Integrable nonlinear maps such as the McMillan map
\cite{mcmillan1971}, whose invariants are known exactly, provide a
controlled setting in which such models can be tested.

Perturbative normal-form methods \cite{forest1998beam,forest1989normal}
describe weakly nonlinear motion accurately, but as series expansions
they lose accuracy at large amplitude, near resonances and close to
the dynamic aperture. Element-by-element tracking is accurate but costly when many
particles must be followed over many turns. Data-driven models can
learn the one-turn map directly, but generic networks do not preserve
the symplectic structure of the dynamics, so their errors can grow over
long-term tracking. Symplectic neural networks \cite{jin2020sympnets} and Hamiltonian neural networks \cite{greydanus2019hamiltonian} remove this problem. On their own, however, they do not explicitly represent the invariants of the motion, so small errors in the learned map can still distort orbits and accumulate phase error, particularly under strong ADTS. A natural remedy is to learn coordinates in which the motion becomes simple. SympNets have been combined with Koopman operator methods, which seek coordinates in which the dynamics is linear [7]. Here we instead combine them with normal-form theory, which lets the turning angle depend on amplitude, as described below.

The normal-form framework \cite{forest1998beam,chao2013handbook}
describes nonlinear dynamics through a nonlinear coordinate
transformation to a simpler representation: away from resonances, a
rotation of each mode whose phase advance depends on amplitude.
Constructing this transformation analytically is difficult, since the
underlying series may converge poorly at large amplitude, and near
resonances the motion cannot be reduced to independent rotations.

In this paper, we learn the normal-form transformation directly from
tracking data. Specifically, we
\begin{enumerate}
  \item use a symplectic neural network (SympNet) \cite{jin2020sympnets}
        to represent the transformation to normal-form coordinates,
        combined with a separate network that gives the
        amplitude-dependent phase advance of each mode;
  \item study a simplified four-dimensional McMillan-type map
        \cite{mcmillan1971}, an integrable nonlinear map whose
        invariants are known exactly, as a controlled test case;

       \item present results on the accuracy of the learned one-turn map, the
      constancy of the learned phase advances, and the conservation of
      the normal-form actions, and discuss the limitations of the model
      at large amplitude. 
  
\end{enumerate}

\section{Model Setup}

Conventional accelerator modeling methods mainly rely on analytical perturbation theory and numerical iterative mapping algorithms. Analytical normal-form methods suffer from limited approximation accuracy under strong nonlinearity and large-amplitude initial conditions, failing to fully characterize distorted orbital features. Traditional numerical mapping strategies can require many repeated computations, especially when many particles and many turns must be tracked. In recent years, data-driven surrogate modeling has emerged as an efficient alternative for physical system prediction, demonstrating outstanding capabilities in fitting complex nonlinear dynamic systems. However, general neural networks lack physical constraint awareness, often destroying the intrinsic symplectic conservation property of Hamiltonian systems. Such unconstrained models typically produce non-physical phase-space volume changes, unstable tune prediction, and serious cumulative drift in long-term autonomous rollout, which severely limits their practical application in accelerator studies.

Existing symplectic neural networks \cite{jin2020sympnets} and
Hamiltonian neural networks \cite{greydanus2019hamiltonian} preserve
the symplectic structure of the dynamics, but they do not explicitly
represent its invariants. Small errors in a learned one-turn map can
therefore still distort orbits and accumulate phase error over many
turns, particularly for large-amplitude orbits with strong
amplitude-dependent tune shift (ADTS). Here we instead build the
normal-form structure into the model \cite{forest1998beam}: the
dynamics are represented as a rotation in normal-form coordinates, so
that the action of each mode is conserved exactly, and the tune
appears explicitly as the phase advance of this rotation.

The goal is to learn the one-turn map $\mathcal M$ of a nonlinear
system in a form that exposes its normal-form structure. Given the
phase-space vector $\boldsymbol z = (x,p_x,y,p_y)^T$, the model
represents the map as
\begin{equation}
\boldsymbol z' = \mathcal M(\boldsymbol z)
= \boldsymbol U\Big(R(\boldsymbol z)\,\boldsymbol U^{-1}(\boldsymbol z)\Big),
\label{eq:map}
\end{equation}
where $\boldsymbol U^{-1}$ is a learnable symplectic transformation from
the physical phase space to the latent normal-form coordinates
$\boldsymbol w = \boldsymbol U^{-1}(\boldsymbol z) = (u_x,v_x,u_y,v_y)$,
$R(\boldsymbol z)$ is a rotation acting in the latent space, and
$\boldsymbol U$ transforms the rotated coordinates back to physical
phase space. The rotation is block-diagonal, one rotation per mode,
\begin{equation}
R=\begin{pmatrix} R(\mu_x) & 0\\ 0 & R(\mu_y)\end{pmatrix},\qquad
R(\mu)=\begin{pmatrix}\cos\mu & \sin\mu\\ -\sin\mu & \cos\mu\end{pmatrix},
\label{eq:rotation}
\end{equation}
with phase advances $\mu_x$, $\mu_y$ (tunes $\nu_{x,y}=\mu_{x,y}/2\pi$)
predicted for each particle by a separate network,  ---- FN(described below).
Unlike conventional analytical normal-form methods, where the
transformation is constructed as a power series, $\boldsymbol U$ and
$\boldsymbol U^{-1}$ here are learned from data. For multi-turn
propagation, the map is applied recursively,
$\boldsymbol z_{n+1}=\mathcal M(\boldsymbol z_n)$, with each output used
as the input of the next turn, as in conventional particle tracking.

\paragraph{FN}is a fully connected network that maps the phase-space state
$s=(x,p_x,y,p_y)$ to two phase advances. It has $L$ linear layers
(default $L=4$, width $h=128$) with $\tanh$ activations between them:
\[
h^{(1)}=\tanh\!\big(W^{(1)}s+b^{(1)}\big),\quad
h^{(\ell)}=\tanh\!\big(W^{(\ell)}h^{(\ell-1)}+b^{(\ell)}\big),\quad
(a,b)^\top = W^{(L)}h^{(L-1)}+b^{(L)} .
\]
The outputs are bounded to $(-\pi,\pi)$:
\[
\mu_x=\pi\tanh a,\qquad \mu_y=\pi\tanh b ,
\]
and define the rotation of Eq(2).
Each block is a rotation, so $R^\top J R = J$ holds exactly for any
$\mu_x,\mu_y$; no additional constraint is needed.

\paragraph{Constraints on the latent rotation.}
The network FN maps the phase-space state to the phase advances
$(\mu_x,\mu_y)$ of the two normal modes. Its outputs are bounded to
$(-\pi,\pi)$ and define a block-diagonal rotation in the latent space,
one rotation per mode. Because each mode is rotated rather than
transformed arbitrarily, its action is conserved exactly, and orbits
remain circles in normal-form coordinates \cite{forest1998beam}. The
phase advances are further constrained by the phase-advance loss to be
constant along each trajectory, so that they depend only on the
invariants while still varying with amplitude, as expected for an
integrable map \cite{mcmillan1971}.


\paragraph{Symplectic Neural Network Architecture.} The symplectic neural network (SympNet) adopted in this work is constructed based on symplectic coordinate transformations and state-dependent rotation blocks, which inherently preserve the symplectic geometry of the phase space. The architecture consists of three core modules: a symplectic encoder $\boldsymbol U^{-1}$, a nonlinear rotation block $\mathcal{R}$, and a symplectic decoder $\boldsymbol U$.

The symplectic encoder $\boldsymbol U^{-1}$ is implemented via a sequence of symplectic affine transformations and separable non-identity activation layers. It maps the physical phase-space coordinates $\boldsymbol z=(x,p_x,y,p_y)$ into a latent space where the transverse oscillations are approximately decoupled. This latent representation separates the amplitude-dependent tune effects from linear phase advance, enabling independent handling of horizontal and vertical motion.

Inside the rotation block $\mathcal{R}$, a multilayer perceptron (FN)
takes the physical phase-space state $(x,p_x,y,p_y)$ as input and
outputs the phase advances $\mu_x$ and $\mu_y$ of the two latent modes,
i.e.\ the tunes $\nu_{x,y}=\mu_{x,y}/2\pi$. These define a
block-diagonal rotation acting on the latent phase-space variables.
Unlike the fixed linear rotation in conventional tracking codes, the
rotation angles vary with the particle state, which captures nonlinear
detuning such as the amplitude-dependent tune shift. A loss term keeps
the phase advances constant along each trajectory, so that in practice
they depend only on the invariants of the motion.

The symplectic decoder $\boldsymbol U$ is the inverse transformation of
$\boldsymbol U^{-1}$. It converts the rotated latent coordinates back
to physical phase space $\boldsymbol z$. Both encoder and decoder are
built as invertible symplectic maps, so the full composition
$\boldsymbol U\circ\mathcal{R}\circ\boldsymbol U^{-1}$ is symplectic
for fixed phase advances; because the phase advances depend on the
state, it is approximately symplectic overall. The entire forward pass
is an explicit deterministic map with no hidden iterative solvers.

\paragraph{Known limitations of the architecture.}
Because the rotation is block-diagonal, the actions $J_x$ and $J_y$ of
the two latent modes are each conserved separately. The model can
therefore only represent dynamics that possess two separately conserved
actions in some set of coordinates; any coupling that transfers action
between the learned modes cannot be represented. In addition, FN computes the phase advances from the full phase-space
state rather than from the actions alone, so the learned map is
symplectic only approximately; computing both phase advances as
derivatives of a single learned function $H(J_x,J_y)$ would make it
exactly symplectic.

\paragraph{Key note}: $\boldsymbol U$ and $\boldsymbol U^{-1}$ are learnable symplectic neural networks rather than constant matrices. This distinguishes the proposed framework from conventional analytical normal-form methods, where the transformation is constructed via power-series expansions.

\begin{figure}[htbp]
\centering
\begin{tikzpicture}[
  box/.style={draw, rounded corners=2pt, align=center, minimum height=1.1cm,
              minimum width=2.6cm, font=\small},
  arr/.style={-{Stealth[length=2mm]}, thick},
  lab/.style={font=\small}
]
\node[lab] (zin) {$\boldsymbol z_n$};
\node[box, right=0.7cm of zin] (enc) {$\boldsymbol U^{-1}$\\ SympNet encoder};
\node[box, right=1.2cm of enc] (rot) {$R(\mu_x,\mu_y)$\\ latent rotation};
\node[box, right=1.2cm of rot] (dec) {$\boldsymbol U$\\ SympNet decoder};
\node[lab, right=0.7cm of dec] (zout) {$\boldsymbol z_{n+1}$};
\node[box, below=1.1cm of rot] (fn) {FN\\ phase advances};
\draw[arr] (zin) -- (enc);
\draw[arr] (enc) -- node[above, lab] {$\boldsymbol w_n$} (rot);
\draw[arr] (rot) -- node[above, lab] {$\boldsymbol w_n'$} (dec);
\draw[arr] (dec) -- (zout);
\draw[arr] (zin) |- (fn);
\draw[arr] (fn) -- node[right, lab] {$\mu_x,\mu_y$} (rot);
\end{tikzpicture}
\caption{Architecture of the model. The SympNet encoder $\boldsymbol U^{-1}$
maps the physical state $\boldsymbol z_n$ to latent normal-form
coordinates $\boldsymbol w_n$, which are rotated by phase advances
$(\mu_x,\mu_y)$ computed by FN from $\boldsymbol z_n$, and mapped back
to physical phase space by the decoder $\boldsymbol U$.}
\label{fig:symp_nf_arch}
\end{figure}
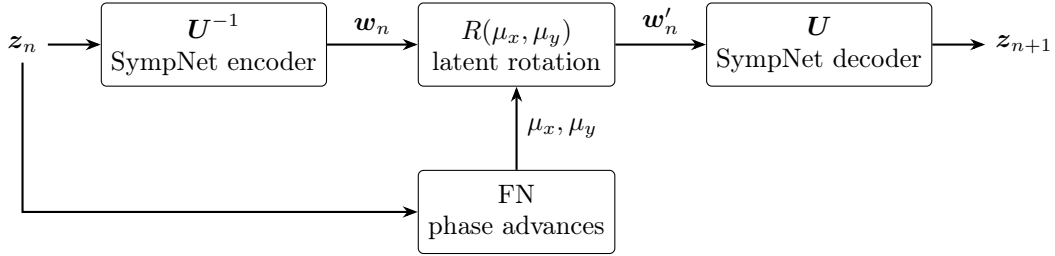

\paragraph{Data Feeding Strategy}

The initial conditions are sampled uniformly in amplitude (not in action)
in each of the $(x,p_x)$ and $(y,p_y)$ planes, with random phases, so
 the data set covers both small-amplitude, nearly linear motion, and
the large-amplitude regime where ADTS is strongest. The training data are generated with a four-dimensional McMillan-type
map. With Twiss parameters $\alpha_x$, $\alpha_y$, $\beta$ and
$\gamma_{x,y}=(1+\alpha_{x,y}^2)/\beta$, one turn reads
\begin{equation}
\begin{aligned}
\bar x &= \alpha_x x + \beta p_x, &
\bar y &= \alpha_y y + \beta p_y,\\
x' &= \bar x, &
y' &= \bar y,\\
p_x' &= -\gamma_x x - \alpha_x p_x + \frac{a\,\bar x}{1+b(\bar x^2+\bar y^2)}, &
p_y' &= -\gamma_y y - \alpha_y p_y + \frac{a\,\bar y}{1+b(\bar x^2+\bar y^2)},
\end{aligned}
\label{eq:mcmillan}
\end{equation}
with $\alpha_x=-0.8$, $\alpha_y=0.5$, $\beta=2$, $a=0.3$ and $b=2$.
Each initial
condition is tracked with Eq.~(\ref{eq:mcmillan}), and consecutive
states are paired as $(\boldsymbol z_k,\boldsymbol z_{k+1})$. No
normalization is applied to the coordinates.

For training, the one-turn pairs are grouped into batches such that
each trajectory lies entirely within one batch, since some of the loss
terms below are computed per trajectory. The order of the batches is
shuffled in every epoch. The training settings are listed in
Table~\ref{tab:training_settings}.

\begin{table}[htbp]
\centering
\caption{Training settings.}
\label{tab:training_settings}
\begin{tabular}{lll}
\hline
Setting & Stage 1 & Stage 2 \\
\hline
Epochs & 120 & 60 \\
Initial learning rate & $5\times10^{-5}$ & $1\times10^{-5}$ (fresh optimizer) \\
Conservation weight $\lambda_\mu$ & 10 & 10 \\
Latent weights $\lambda_{l,x}, \lambda_{l,y}$ & 2 & 2 \\
Radius  weights $\lambda_{J,x}, \lambda_{J,y}$ & 0.2, ramped over 25 epochs & 0.6, no ramp \\
Plateau patience & 15 epochs & 10 epochs \\
\hline
\multicolumn{3}{l}{Shared settings} \\
\hline
Training data & \multicolumn{2}{l}{200 trajectories $\times$ 400 turns} \\
Initial amplitudes & \multicolumn{2}{l}{uniform in $[0.05,\,0.70]$ per plane, random phases} \\
Batch size & \multicolumn{2}{l}{about 2048 pairs, whole trajectories} \\
Optimizer & \multicolumn{2}{l}{AdamW, weight decay $10^{-6}$, gradient clipping at norm 1} \\
Learning-rate schedule & \multicolumn{2}{l}{halved on plateau, minimum $10^{-6}$} \\
SympNet & \multicolumn{2}{l}{6 blocks of 6 activation layers, 5 linear mixers} \\
$U_{\text{rough}}$ fit & \multicolumn{2}{l}{10\% of points closest to the origin} \\
FN & \multicolumn{2}{l}{$L=4$ linear layers of width $h=128$} \\
Precision, seed & \multicolumn{2}{l}{float64, random seed 1234} \\
\hline
\end{tabular}
\end{table}

\paragraph{Loss-function} 
The phase advances $(\mu_x,\mu_y)$ are internal quantities, not
targets: there are no ground-truth labels for them. What is available
is data, i.e.\ observed one-turn pairs $(\boldsymbol z_k,\boldsymbol
z_{k+1})$, and the model is trained so that the resulting map
reproduces these dynamics. The total loss is a weighted sum of terms,
all computed on one-turn pairs. Write
$\boldsymbol w_k=\boldsymbol U^{-1}(\boldsymbol z_k)$ for the latent
coordinates of a data point, $R_k$ for the rotation with the phase
advances FN assigns to $\boldsymbol z_k$, and $N$ for the number of
pairs in a batch. The augmented loss terms consist of latent $x$ and $y$ losses, radius $x$ and $y$ losses, the predicted loss, and the combined total loss. These terms encourage the learned model to preserve the observed dynamical structure across different amplitudes.

The latent $x$ and $y$ loss is
\begin{equation}
\mathcal L_{\text{latent}} = \lambda_x \frac{1}{N}\sum_{k=1}^N (x_k-\hat x_k)^2 + \lambda_y \frac{1}{N}\sum_{k=1}^N (y_k-\hat y_k)^2,
\end{equation}
where $\lambda_x,\lambda_y$ are weighting coefficients for the latent $x$ and $y$ coordinate errors, and $N$ is the number of samples.


For each transverse mode, define the latent actions of an encoded data
point $\boldsymbol w_k=\boldsymbol U^{-1}(\boldsymbol z_k)=(u_{x,k},v_{x,k},u_{y,k},v_{y,k})$,
\[
J_{x,k}=u_{x,k}^2+v_{x,k}^2,\qquad
J_{y,k}=u_{y,k}^2+v_{y,k}^2 .
\]
For a trajectory $m$ with points $k\in m$, let
$\bar J_{x,m}$ and $\operatorname{Var}_m(J_x)$ be the mean and variance
of $J_{x,k}$ along that trajectory, and likewise for $J_y$. The radius
loss reads
\begin{equation}
\mathcal L_{\text{radius}}
= \frac{\lambda_{r,x}}{M}\sum_{m=1}^{M}\frac{\operatorname{Var}_m(J_x)}{\bar J_{x,m}^{\,2}}
+ \frac{\lambda_{r,y}}{M}\sum_{m=1}^{M}\frac{\operatorname{Var}_m(J_y)}{\bar J_{y,m}^{\,2}} ,
\end{equation}
where $M$ is the number of trajectories in a batch, and $\lambda_{r,x}$ and
$\lambda_{r,y}$ weight the horizontal and vertical modes. The loss vanishes
when every observed orbit is a circle in the learned coordinates.
Normalizing by the mean action gives small and large orbits equal
weight, and no model prediction is involved: the loss acts on the
encoded data only.


Each training sample consists of an initial value
$(x, p_x, y, p_y)$ and its image $M(x, p_x, y, p_y)$ under the map.
The network $\mathcal N$ is trained to reproduce this image, by
minimizing
\begin{equation}
\mathcal L_{\text{pred}} = \frac{1}{N}\sum_{k=1}^{N}
\big\| M(x_k, p_{x,k}, y_k, p_{y,k})
      - \mathcal N(x_k, p_{x,k}, y_k, p_{y,k}) \big\|^2 ,
\end{equation}
where $N$ is the number of training samples.

\paragraph{Conservation loss.}
For each point $\boldsymbol z_k$, FN assigns the phase advances
$\big(\mu_{x,k},\mu_{y,k}\big)=\mathrm{FN}(\boldsymbol z_k)$. For a
trajectory $m$ with $N_m$ points in the batch, let
\[
\bar\mu_{x,m}=\frac{1}{N_m}\sum_{k\in m}\mu_{x,k},\qquad
\operatorname{Var}_m(\mu_x)=\frac{1}{N_m}\sum_{k\in m}\big(\mu_{x,k}-\bar\mu_{x,m}\big)^2 ,
\]
and likewise for $\mu_y$. The conservation loss is
\begin{equation}
\mathcal L_\text{cons} = \frac{1}{M}\sum_{m=1}^{M}
\Big[\operatorname{Var}_m(\mu_x) + \operatorname{Var}_m(\mu_y)\Big],
\label{eq:phase_loss}
\end{equation}
averaged over the $M$ trajectories in the batch. It vanishes when the
phase advances are constant along every trajectory, while still
allowing different trajectories to have different phase advances.


The total loss combines all components:
\begin{equation}
\mathcal L_{\text{total}} = \lambda_{\text{latent}} \mathcal L_{\text{latent}} + \lambda_{\text{radius}} \mathcal L_{\text{radius}} + \lambda_{\text{pred}} \mathcal L_{\text{pred}}
+\lambda_{\text{cons}} \mathcal L_{\text{cons}},
\end{equation}
where $\lambda_{\text{latent}},\lambda_{\text{radius}},\lambda_{\text{pred}},\lambda_{\text{cons}},$ are hyperparameters controlling the relative contribution of each loss component.


\section{Results}

The trained model learns an approximate normal-form transformation
for the nonlinear symplectic map~$\mathcal M$: a coordinate change
\[
\boldsymbol U^{-1}:\ (x, p_x, y, p_y) \mapsto (u_x, v_x, u_y, v_y)
\]
under which the motion, visibly non-circular and amplitude-dependent in
the original variables, becomes close to a rotation in the learned
variables,
\[
\boldsymbol U^{-1} \circ \mathcal M \circ \boldsymbol U \;\approx\;
R(\mu_x,\mu_y),
\]
with $R(\mu_x,\mu_y)$ the block-diagonal rotation of
Eq.~(\ref{eq:rotation}). The phase advances $(\mu_x,\mu_y)$ assigned by
FN stay nearly constant along a given trajectory, with standard
deviations of $1$--$4\times10^{-5}$~rad over the trajectories examined.
This is the defining property of action-angle coordinates of an
integrable system: the angle advances at a constant rate turn by turn,
while the action is conserved. The quality of the learned normal form
can be quantified by the residual
\[
\epsilon_{\mathrm{NF}}
= \Big\langle \big\| R(\boldsymbol z_k)\,\boldsymbol U^{-1}(\boldsymbol z_k)
- \boldsymbol U^{-1}(\boldsymbol z_{k+1}) \big\|^2 \Big\rangle ,
\]
the mean squared difference, per latent coordinate, between the rotated
latent point and the latent point of the true next state.

\begin{figure}[htbp]
\centering
\includegraphics[width=0.75\textwidth]{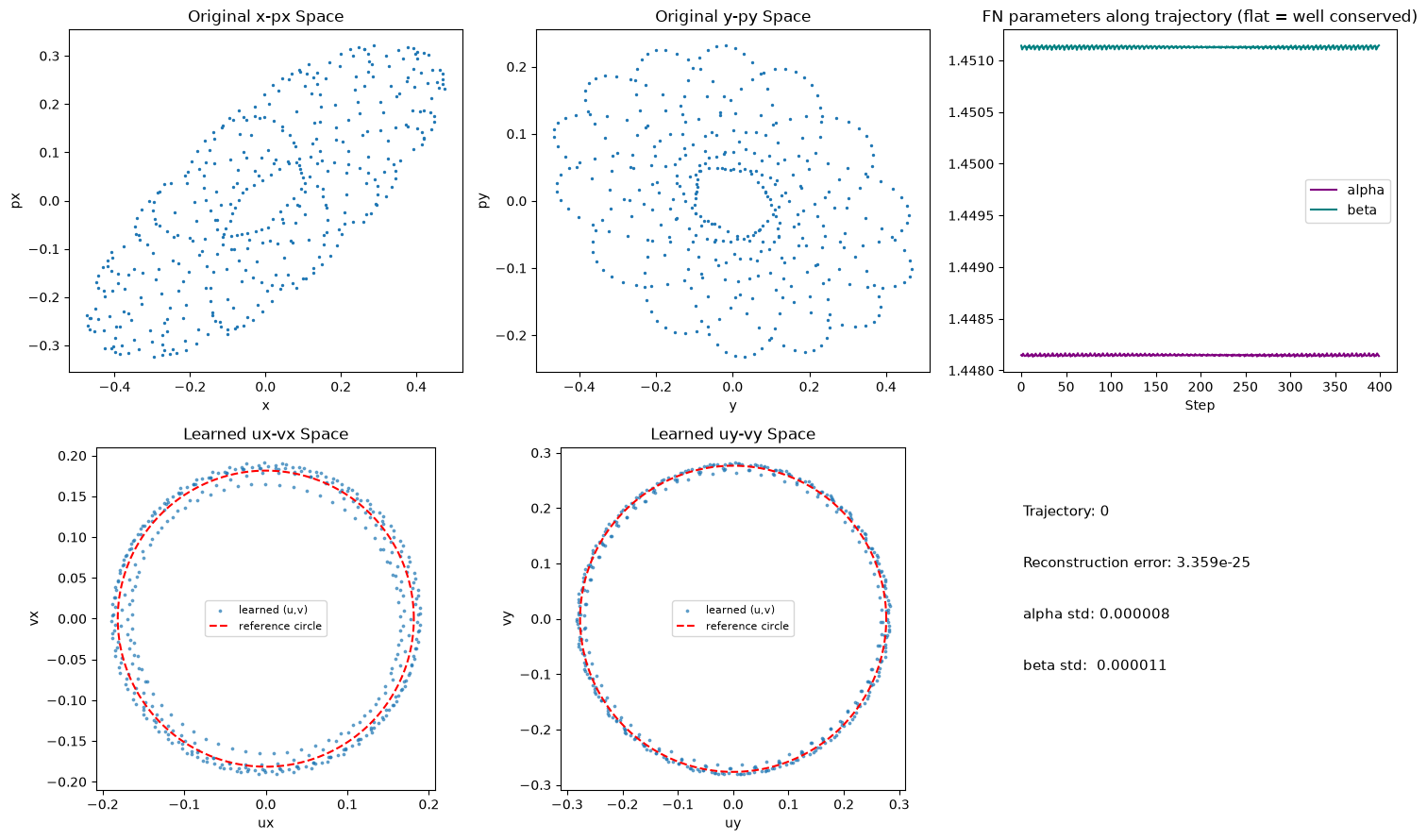}
\caption{Phase-space visualization and quantitative metrics for a single small-amplitude particle trajectory.}
\label{fig:phase1}
\end{figure}

\begin{figure}[htbp]
\centering
\includegraphics[width=0.7\textwidth]{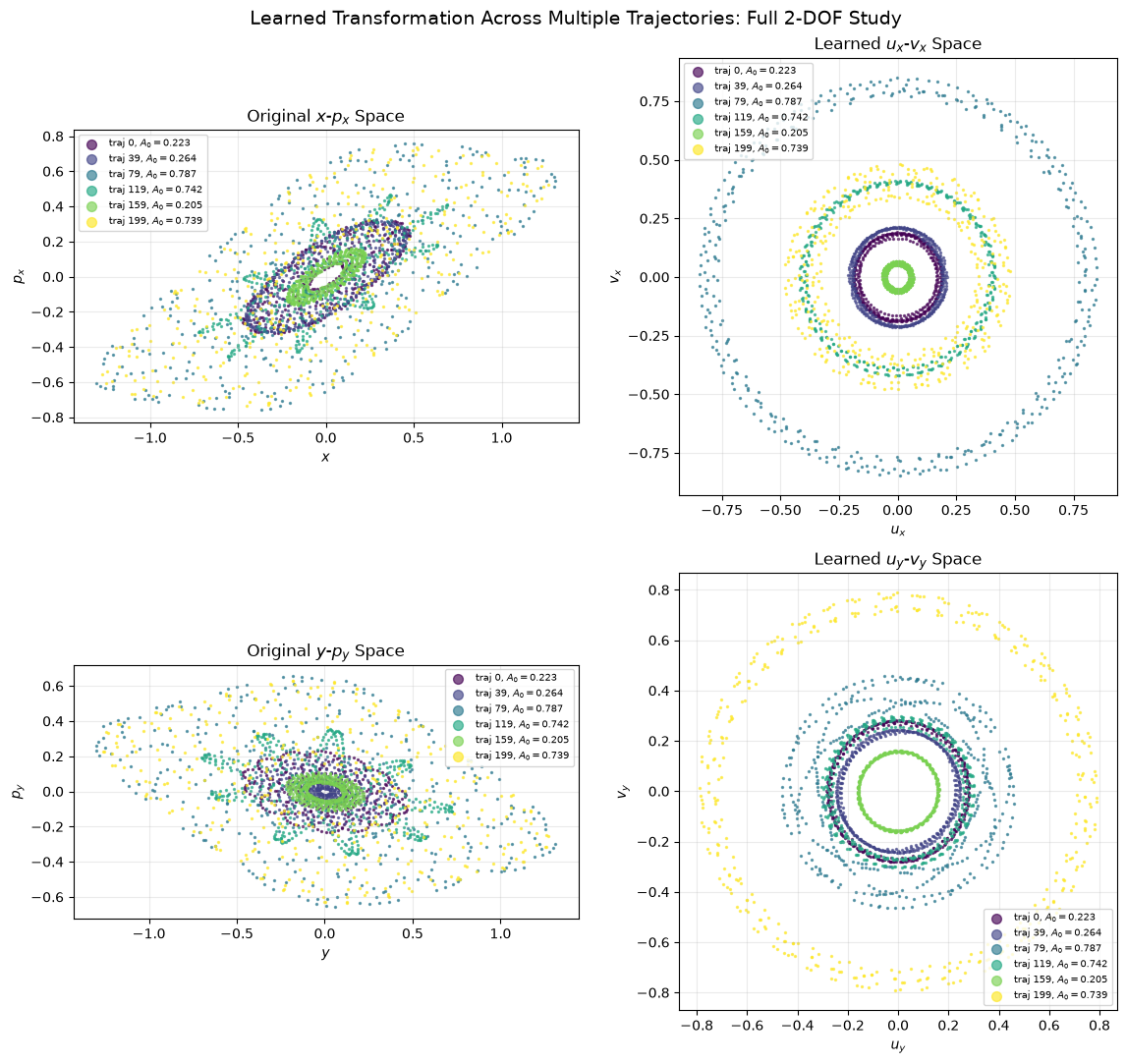}
\caption{Phase-space portraits of multiple particle trajectories with different oscillation amplitudes.}
\label{fig:phase2}
\end{figure}

Figures~\ref{fig:phase1} and \ref{fig:phase2} show that the learned
transformation maps the non-circular orbits of the McMillan-type map to
nearly circular orbits in normal-form coordinates. The phase advances
$(\mu_x,\mu_y)$ produced by FN remain nearly constant along each
trajectory, with standard deviations of order $10^{-5}$~rad, consistent
with an integrable map. The SympNet transformation is symplectic by
construction, and the rotation is symplectic for fixed phase advances;
since the phase advances depend on the state, the full map is
symplectic only approximately. At large amplitude, the orbits show
noticeable radial dispersion in normal-form coordinates, which grows
gradually with amplitude; its cause has not yet been established.

\begin{figure}[htbp]
\centering
\includegraphics[width=0.75\textwidth]{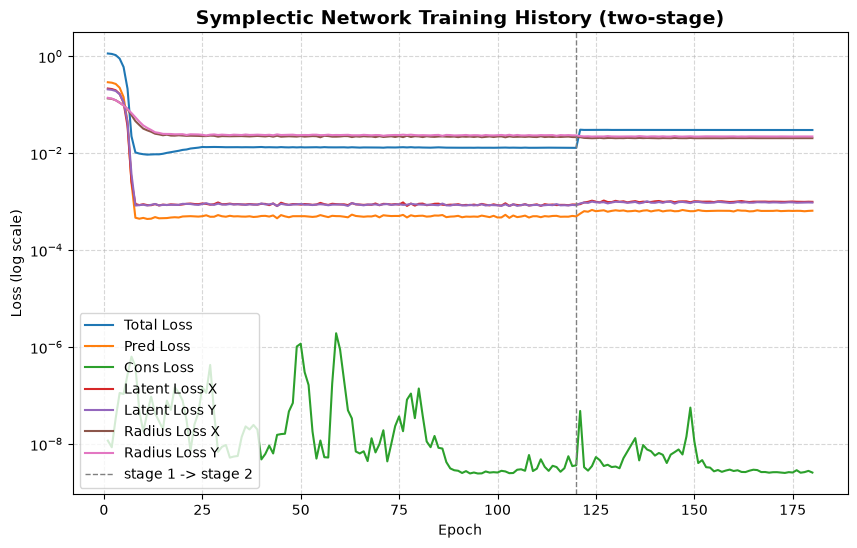}
\caption{Loss function during training history.}
\label{fig:loss}
\end{figure}

Figures~\ref{fig:phase1} and \ref{fig:phase2} indicate that the model
retrieved an approximate normal-form (action-angle) coordinate system
for this nonlinear symplectic map at small and moderate amplitude; at
the largest amplitudes in the training data, the learned orbits are
noticeably less circular.

Figure~\ref{fig:phase1} shows this directly: the rotation parameters
$(\alpha, \beta)$ remain essentially flat across a full trajectory
($\operatorname{std}(\alpha) \sim 10^{-4}$), and the trajectory traced in the
learned $(u_x, v_x, u_y, v_y)$ space sits close to the reference circle.
This is exactly the defining signature of action-angle coordinates: the angle
advances at a constant rate turn by turn, and the action (amplitude) is
conserved --- which is not true of the same trajectory in the original
$(x, p_x, y, p_y)$ variables, where the motion is visibly non-circular.
(The near machine-precision reconstruction error is a consequence of the
exact invertibility of the symplectic architecture and serves as a
consistency check rather than evidence of the normal form.)

Figure~\ref{fig:phase2} shows where this result holds and where
it does not: at small-to-moderate initial amplitudes, the learned
$(u_x, v_x)$ contours are noticeably closer to circular than the original
phase-space contours. At the largest tested amplitude, however, the learned
contour remains distorted and non-convex, meaning the model has not (yet)
demonstrated that it has found the true normal-form coordinates there. This
may be a genuine physical limit of the system --- a real nonlinear
lattice/map typically stops being close to integrable past some amplitude
(the ``dynamic aperture'' idea) --- or a training/data-coverage limitation;
the two have not yet been distinguished.

Taken together, the network has learned to discover numerically the
invariant structure of the dynamics, not merely to fit input--output
pairs: it found a coordinate system in which this nonlinear problem
looks simple. This claim is well supported at small and moderate
amplitude and remains open at large amplitude. The degradation at large
amplitude may reflect a limitation of the block-diagonal
representation, or limited coverage of large amplitudes in the training
data; these two causes have not yet been distinguished.

\section{Conclusions and Discussion}

We have presented a model that learns the normal-form transformation of
a nonlinear symplectic map from tracking data. A SympNet provides an
exactly invertible, symplectic transformation to latent coordinates, in
which the dynamics are represented as a rotation of each mode with
amplitude-dependent phase advances given by FN network. Training uses one-turn pairs drawn from tracked trajectories. The loss
weights follow a two-stage schedule that first fits the dynamics and
then tightens action conservation. In the first stage, the prediction,
latent and phase-advance terms are active from the start, while the
action weights are ramped up gradually, so that early training
captures the amplitude-dependent dynamics. The second stage continues
from the trained networks with a fresh optimizer, a lower learning
rate and larger action weights. The resulting model gives stable tune
predictions and well-reproduced orbit shapes for small- and
medium-amplitude particles (Fig.~\ref{fig:phase2}); its symplectic structure
comes from the SympNet architecture rather than from the loss function.


The proposed method has several limitations. First, large-amplitude
orbits show residual oscillations in normal-form coordinates, and
increasing the weight of the action loss reduces it only slightly. It
is not yet established whether this reflects the block-diagonal form of
the rotation or limited training data at large amplitude. Second,
because FN computes the phase advances from the full phase-space state
rather than from the actions alone, the learned map is symplectic only
approximately. Finally, the method has so far been demonstrated only on
a McMillan-type map; its application to realistic lattices remains to
be tested.

Future work will focus on addressing the above deficiencies to further improve model generalization and physical robustness. Redesigning the loss function architecture is essential to maintain stable and effective symplectic regularization throughout the training process. Moreover, improved latent-space modeling strategies will be explored to eliminate amplitude-dependent systematic bias, enabling high-fidelity beam simulation across full phase-space regions. In summary, the proposed framework provides an efficient and physically reliable surrogate solution for fast nonlinear beam dynamics simulation, offering great potential for future high-precision accelerator lattice optimization and long-term beam orbit prediction.


\end{document}